\documentclass[aps,reprint,prd,superscriptaddress,nofootinbib,amsmath,amssymb]{revtex4-2}

\usepackage{xcolor}
\usepackage{mathrsfs}
\usepackage[colorlinks=true,citecolor=blue]{hyperref}
\usepackage{orcidlink}

\begin{document}

\title{Non-Noetherian conformal Cheshire effect}

\author{Eloy Ay\'on-Beato\,\orcidlink{0000-0002-4498-3147}}
\email{eloy.ayon-beato-at-cinvestav.mx} 
\affiliation{Departamento de F\'{\i}sica, CINVESTAV-IPN, A.P.~14740, C.P.~07360, Ciudad de M\'exico, M\'exico}

\author{Mokhtar Hassaine\,\orcidlink{0009-0003-3159-5916}}
\email{hassaine-at-inst-mat.utalca.cl} 
\affiliation{Instituto de Matem\'atica, Universidad de Talca, Casilla 747, Talca, Chile}

\author{Pedro A. S\'anchez\,\orcidlink{0009-0008-7437-0561}}
\email{pedro.sanchez.s-at-cinvestav.mx} 
\affiliation{Departamento de F\'{\i}sica, CINVESTAV-IPN, A.P.~14740, C.P.~07360, Ciudad de M\'exico, M\'exico}

\begin{abstract}
The gravitational Cheshire effect refers to the possibility of turning off the gravitational field while still leaving an imprint of the nonminimal coupling of matter to gravity. This allows nontrivial solutions in flat spacetime for which no backreaction is possible. The effect was originally shown to manifest itself for standard nonminimal couplings, such as those allowing conventional conformally invariant scalar fields. Recently, the most general scalar field action yielding a conformally invariant second-order equation was constructed, and entails a more involved nonminimal coupling explicitly breaking the conformal invariance of the action without spoiling it in the equation. We have succeeded in fully describing the spherically symmetric stealth solutions on flat spacetime supporting the Cheshire effect within this general non-Noetherian conformal theory. The allowed configurations are divided into two branches: The first one essentially corresponds to an extension of the solutions already known for the standard Noetherian conformal theory. The second branch is only possible due to the non-Noetherian conformal contribution of the action. The complete characterization of this branch is expressed by a nonlinear first-order partial differential equation. We have found the general solution of this equation using both seemingly new and well-established mathematical tools.
\end{abstract}

\maketitle

%%%%%%%%%%%%%%%%%%%%%%%%%%%%%%%%%%%%%%%%%%%%%%%%%%%%%%%%%%
\section{Introduction \label{sec:intro}} 
%%%%%%%%%%%%%%%%%%%%%%%%%%%%%%%%%%%%%%%%%%%%%%%%%%%%%%%%%%

It is evident that nonminimal couplings to gravity, arising due to the explicit presence of curvature in matter Lagrangians, vanish in the flat spacetime limit. However, it is less expected that after taking such a limit some nonminimal contributions to the variation persist, leaving an imprint on the energy-momentum tensor. This phenomenon has been dubbed the \emph{gravitational Cheshire effect} \cite{Ayon-Beato:2005yoq}, making a reference to the Cheshire cat in Lewis Carroll's \emph{Alice's Adventures in Wonderland}, which disappears leaving behind only its grin. The effect is manifested through the existence of nontrivial solutions where the nonminimal and minimal contributions cancel each other, leading to a vanishing energy-momentum tensor in flat spacetime. In fact, these are examples of \emph{gravitational stealth} configurations whose definition extends beyond flat spacetime, being genuine nontrivial solutions to Einstein's equations by simultaneously canceling both their gravitational and matter contributions \cite{Ayon-Beato:2004nzi}. 

The Cheshire effect has been exhibited in particular for standard conformal scalar fields \cite{Callan:1970ze} whose action is 
\begin{subequations}\label{eq:stdCSF} {\small
\begin{equation}\label{eq:stdCSF_action}
S_\text{s}[\Phi,g] = - \frac12\!\int\!d^4 x \sqrt{-g} \!\left(\!  \nabla_\mu\Phi\nabla^\mu\Phi + \frac16 R \Phi^2 + 2\lambda \Phi^4 \!\right)\!,
\end{equation}}%
yielding the scalar field equation 
\begin{equation}\label{eq:stdCSFeom}
\Box\Phi-\frac16R\Phi-4\lambda\Phi^3=0,
\end{equation}
and energy-momentum tensor 
\begin{align} 
T_{\mu\nu}^{\text{s}} ={}& \nabla_\mu \Phi \nabla_\nu \Phi - \left( \frac12 \nabla_\alpha\Phi\nabla^\alpha\Phi + \lambda \Phi^4 \right) g_{\mu \nu} \nonumber\\
& + \frac{1}{6} \left( g_{\mu\nu} \Box-\nabla_{\mu}\nabla_{\nu}+G_{\mu\nu} \right )\Phi^2. \label{eq:T_munu_std}
\end{align}
\end{subequations}
Here, $\lambda$ is the coupling constant of the conformal self-interaction and the symbols $\Box$, $R$, and $G_{\mu\nu}$ stand for the d'Alembertian operator, the scalar curvature, and the Einstein tensor, respectively. The peculiarity of this system is that action \eqref{eq:stdCSF_action} is invariant under the conformal transformations 
\begin{align}\label{eq:CT}
g_{\mu\nu} &\mapsto \Omega^2(x) g_{\mu\nu}, & 
\Phi &\mapsto \Omega^{-1}(x)\Phi.
\end{align}
In general, satisfying such property in the action will imply that the energy-momentum tensor is traceless modulo the equation of motion. To connect with our previous discussion, the last terms related to $\Phi^2$ in \eqref{eq:T_munu_std} are due to the nonminimal coupling in the action, and it is clear that in the flat limit $g_{\mu\nu}\to\eta_{\mu\nu}$ these terms do not vanish. In fact, they compensate with the previous minimal contributions in order to allow the existence of flat spacetime stealths that are the direct manifestation of the conformal Cheshire effect, as shown in \cite{Ayon-Beato:2005yoq}.

To make our terminology clear, the stealths are defined as the nontrivial extrema of any matter action not only with respect to the matter content but also with respect to the gravitational background. The first such example was shown to arise on the static BTZ black hole in three dimensions for a scalar field with the simplest nonminimal coupling to gravity, which contains the conformal coupling \eqref{eq:stdCSF} as a particular case \cite{Ayon-Beato:2004nzi}.  It was later on generalized for the same theory on (A)dS spacetimes of arbitrary dimension \cite{Ayon-Beato:SAdS} or even on more exotic spacetimes, but likewise holographically relevant at non-relativistic limits \cite{Ayon-Beato:2015qfa}. Applications of stealth configurations in the context of cosmology were also reported in Refs.~\cite{Banerjee:2006pr,Maeda:2012tu,Ayon-Beato:2013bsa,Ayon-Beato:2015mxf}.

Scalar-tensor theories, particularly those within the Horndeski class \cite{Horndeski:1974wa}, have proven to be an excellent laboratory for exhibiting stealth configurations. These examples are not limited to zero mass gravitational states but also include configurations defined on black holes, such as the Schwarzschild metric \cite{Babichev:2013cya}, relying on having a constant kinetic term, see also Refs.~\cite{Minamitsuji:2018vuw,Giardino:2023qlu}. More recently, it has been shown that in so-called beyond Horndeski theories \cite{Gleyzes:2014dya}, there exist stealth solutions with a non-constant kinetic term \cite{Bakopoulos:2023fmv}; for a detailed discussion on stealth solutions in the context of such theories, see \cite{Lecoeur:2024kwe}. The existence of such black hole stealths, which violate certain assumptions of the standard ``no-hair'' theorems valid in General Relativity, could prove to be of great use to distinguish such theory from scalar-tensor ones by means of future observations. In this perspective, a stealth on the top of the Kerr black hole was constructed in \cite{Charmousis:2019vnf}, and it turns out that its disformal extension leads to another rotating black hole solution with different physical properties from those characterizing the unique prediction of General Relativity \cite{Anson:2020trg}.

In this article, we extend the known conformal realizations of the gravitational Cheshire effect provided in \cite{Ayon-Beato:2005yoq} by considering the most general action principle for a scalar field whose equation of motion is of second order and conformally invariant. This theory was identified in \cite{Ayon-Beato:2023bzp}, inspired by seminal results of \cite{Fernandes:2021dsb}. The most innovative contribution to the conformally invariant equation is a nonminimal coupling, initially identified in \cite{Fernandes:2021dsb}, whose inclusion breaks the conformal symmetry of the action but not of the equation; such symmetry was dubbed non-Noetherian conformal symmetry in \cite{Ayon-Beato:2023bzp}. Furthermore, it stands out that this new contribution makes the scalar field equation of motion fully non-linear, in contrast with the quasilinear standard conformal equation \eqref{eq:stdCSFeom}.
 
To begin our exploration, it is natural to ask whether the standard conformal stealth configurations on flat spacetime \cite{Ayon-Beato:2005yoq}, retain its stealth behavior after introducing non-Noetherian conformal contributions. This task is straightforward even without imposing any symmetry on the scalar field because these Noetherian conformal configurations obey specific separability properties, as shown in \cite{Ayon-Beato:2005yoq}, which are also effective for the full theory. Conversely, examining the broader stealth spectrum of the full conformal theory beyond this separability presents a major challenge. The new stealth constraints are entirely nonlinear, making the quasilinearity of the standard case exploited in \cite{Ayon-Beato:2005yoq} inapplicable. To advance in this new study of the conformal gravitational Cheshire effect, we are focusing on spherically symmetric configurations that are still time-dependent. Unexpectedly, this simple assumption leads to the mutual cancellation of all second-order nonlinear contributions in the relevant independent constraints. This opens the door to a new branch of the gravitational Cheshire effect beyond separability, encoded in a single nonlinear first-order PDE. Remarkably, this branch exists only for a family of nonvanishing values of the non-Noetherian conformal coupling constant.

The full characterization of the new branch demands the use of sophisticated PDE tools. Indeed, finding the \emph{general solution} of a fully nonlinear first-order PDE is not as straightforward as it is for its linear and quasilinear counterparts, where the method of characteristics is enough. The standard approach in the fully nonlinear case, originally formulated by Lagrange (see \cite{Demidov:1982} and references therein), involves finding the \emph{envelope} of a so-called \emph{complete integral}, subject to the condition that one of its parameters is a general function of the others. This results in an implicit representation of the general solution even in the simplest nonlinear examples.

In contrast to the sophistication of Lagrange's methodology, we introduce a seemingly novel elementary strategy here, leveraging the fact that the involved nonlinear first-order PDE can be cast into a standard Hamilton-Jacobi equation. Despite being a well-known equation in mechanics, finding its general solution is not common since a complete solution suffices to obtain the general integral of the equations of motion of the mechanical system \cite{*[{}] [{. See \S 47.}] Landau:1960}. In this work, we address this by exploiting the quadratic dependence on the derivatives in the obtained Hamilton-Jacobi equation, allowing its identification with the definition of an hyperbola. Imposing integrability conditions on the hyperbola parameterization gives rise to a quasilinear first-order PDE for the local parameter. Naturally, this equation can be integrated using the more conventional method of characteristics.

The paper is organized as follows. In Sec.~\ref{sec:Non-Noeth-CSF}, we briefly present the most general conformally invariant second-order equation arising from an action principle and the related theory. In Sec.~\ref{sec:noeth-like}, we continue by reviewing the previously known separable stealth solutions defined on flat spacetime for the standard conformal case \eqref{eq:stdCSF}. Then, we analyze how these solutions prevail when the most general non-Noetherian conformal theory is considered. In Sec.~\ref{sec:strictly-nonNoeth}, we will show that turning on the non-Noetherian conformal coupling opens the possibility of obtaining radically new spherically symmetric conformal stealths beyond separability. We manage to prove that this new branch only exists for a precise family of nonvanishing values of the related coupling constant, and is exclusively determined by a single nonlinear first-order PDE. The general solution of the latter will be found using both the seemingly new and also the well-established PDE methods outlined above. Hence, we will provide several representations that can be useful for different applications and limits. Concretely, using the aforementioned quasilinearization we found three different representations for the general solution in Subsec.~\ref{subsec:HJq}, two of them are implicit as expected. Notably, coordinate freedom also provides an explicit representation in a special system adapted to the problem, which is unfortunately awkward for expressing the Minkowski metric. We also apply the standard PDE approach relying on the Lagrange technique. When used on the complete integral derived from the so-called Lagrange-Charpit method, one of the representations obtained by quasilinearization is recovered in Subsec.~\ref{subsec:GSfromLC}. Since separable solutions are also a source of complete integrals we address them in Subsec.~\ref{subsec:GSnonNfromN}. There, we also get the unexpected result that the previously mentioned extension of the separable Noetherian conformal stealths of \cite{Ayon-Beato:2005yoq} can be also used as seed in the Lagrange approach to derive the most general non-Noetherian conformal stealths. Supplementary to this, in Subsec.~\ref{subsec:S&Ps},  we also present the so-called \emph{singular solution} not contained in the general one, and show how several particular cases emerge from the general solution. The last section is devoted to a discussion concerning our results. To be as instructive as possible, we briefly review in App.~\ref{app:nonlinear-PDEs} all the PDE definitions and methods in which the main results are founded. Explicit expressions for relevant conformally invariant scalars are provided in App.~\ref{app:R_GBtilde}.

%%%%%%%%%%%%%%%%%%%%%%%%%%%%%%%%%%%%%%%%%%%%%%%%%%%%%%%%%%
\section{Non-Noetherian conformal scalar fields
\label{sec:Non-Noeth-CSF}} 
%%%%%%%%%%%%%%%%%%%%%%%%%%%%%%%%%%%%%%%%%%%%%%%%%%%%%%%%%%

The most general second-order scalar field equation invariant upon a conformal transformation \eqref{eq:CT} and arising from an action principle in four dimensions can be compactly written as \cite{Ayon-Beato:2023bzp}
\begin{equation}\label{eq:nonNoethCSFeom}
\Box\Phi-\frac16R\Phi-\Phi^3\left(4\lambda + \frac{\alpha}{2} \tilde{\mathscr{G}}
-\tilde{g}^{\alpha\beta}\frac{\partial
f}{\partial\tilde{g}^{\alpha\beta}} +2f\right)=0.
\end{equation}
Here, we denote by
\begin{equation} \label{eq:aux-metric}
\tilde{g}_{\mu\nu}=\Phi^2g_{\mu\nu},
\end{equation}
an auxiliary metric which is conformally invariant by construction \eqref{eq:CT}. Furthermore, $\alpha$ is a coupling constant mediating a new nonminimal coupling to gravity expressed through $\tilde{\mathscr{G}}=\tilde{R}^2-4\tilde{R}_{\alpha\beta}\tilde{R}^{\alpha\beta} +\tilde{R}_{\alpha\beta\mu\nu}\tilde{R}^{\alpha\beta\mu\nu}$, which is nothing but the Gauss-Bonnet scalar built from the auxiliary metric \eqref{eq:aux-metric}, and whose explicit expression is given in Eq.~\eqref{eq:GBtilde}. Additionally,  $f=f(\tilde{g}_{\mu\nu},\tilde{C}^\alpha_{~\beta\mu\nu})$ is an arbitrary function of the auxiliary Weyl tensor, only subject to the condition
\begin{equation} \label{eq:f-Weyl_init-cond}
f(\tilde{g},0) = 0.
\end{equation}
Since the new ingredients in Eq.~\eqref{eq:nonNoethCSFeom} are all constructed from the auxiliary metric $\tilde{g}_{\mu\nu}$, it is clear that this equation is again conformally invariant. We remark that \eqref{eq:nonNoethCSFeom} is a fully nonlinear equation since $\tilde{\mathscr{G}}$ is quadratic on the second-order highest derivatives of $\Phi$, see \eqref{eq:GBtilde}, and only for $\alpha=0$ this equation becomes quasilinear by involving exclusively linear contributions of the derivatives of $\Phi$.

As was also shown in \cite{Ayon-Beato:2023bzp}, the
action yielding to  \eqref{eq:nonNoethCSFeom} reads
\begin{subequations}\label{eq:Non-NoethCSFaction} {\small
\begin{align} 
S[\Phi,g] ={}& S_\text{s}[\Phi,g] + S_\text{nN}[\Phi,g] \nonumber \\
&- \frac12 \!\int\! d^4 x \sqrt{-g} \,\Phi^4f(\Phi^2g_{\mu\nu},C^\alpha_{~\beta\mu\nu}),
\end{align}}%
where $S_\text{s} [\Phi,g]$ was given in \eqref{eq:stdCSF_action} and \cite{Fernandes:2021dsb} {\small
\begin{align}
S_\text{nN}[\Phi,g]={}&-\frac\alpha2\!\!\int\!\!d^4 x \sqrt{-g}
\biggl(\ln(\Phi)\mathscr{G} 
-\frac4{\Phi^2}G^{\mu\nu}\nabla_\mu\Phi\nabla_\nu\Phi
\notag \\ 
&-\frac4{\Phi^3}(\nabla_\mu\Phi\nabla^\mu\Phi)\Box\Phi
+ \frac2{\Phi^4}(\nabla_\mu\Phi\nabla^\mu\Phi)^2\biggr).
\label{eq:SnonNoether}
\end{align}}%
\end{subequations}
Here, $C^\alpha_{~\beta\mu\nu}$ and $\mathscr{G}$ refer to the components of the Weyl tensor and the Gauss-Bonnet scalar of the background metric $g_{\mu\nu}$, respectively. Notice that the full action is parity invariant in the scalar field, $\Phi\mapsto-\Phi$. For $\alpha = 0$ and $f = 0$, the action reduces to the standard conformal action \eqref{eq:stdCSF_action}. Interestingly, not only this term is conformally invariant but also the last one defining the Weyl coupling in the full action. In fact, both can be rewritten only in terms of the auxiliary metric up to boundary terms
\begin{equation}
- \frac12\!\int\!d^4 x \sqrt{-\tilde{g}} \!\left(\!   \frac16 \tilde{R} + 2\lambda + f(\tilde{g}_{\mu\nu},\tilde{C}^\alpha_{~\beta\mu\nu}) \!\right)\!+\text{b.t.},
\end{equation}
where $\tilde{R}$ is the scalar curvature of the auxiliary metric $\tilde{g}_{\mu\nu}$, see Eq.~\eqref{eq:Rtilde} for its explicit expression. This is compatible with the result proved in \cite{Ayon-Beato:2023bzp} (see their Eq.~(9) and the related discussion) that any conformally invariant scalar density built from the scalar field and the metric must exclusively be expressed in terms of the auxiliary metric. However, this is no longer the case for the $\alpha$-extension described by action \eqref{eq:SnonNoether}, which when expressed in terms of the auxiliary metric keeps a dependence on the scalar field 
\begin{equation}
S_\text{nN}[\Phi,g]=-S_\text{nN}[\Phi^{-1},\tilde{g}] +\text{b.t.},
\end{equation}
and hence, explicitly breaks the conformal invariance of the action. This was first carefully stressed by Fernandes \cite{Fernandes:2021dsb} using an infinitesimal argument. In summary, the conformal symmetry of Eq.~\eqref{eq:nonNoethCSFeom} does not have a Noetherian origin, hence, such equation describes \emph{non-Noetherian conformal scalar fields} \cite{Ayon-Beato:2023bzp}.\footnote{See Refs.~\cite{Jackiw:2005su,Ayon-Beato:2023lrn} for the two-dimensional non-Noetherian analog.} 

The variation of action \eqref{eq:Non-NoethCSFaction} with respect to the metric gives the energy-momentum tensor whose full expression is reported in Refs.~\cite{Fernandes:2021dsb,Ayon-Beato:2023bzp}. It is more compactly expressed if the scalar field is redefined according to its conformal weight
\begin{equation}\label{eq:sigma}
\sigma = \frac{1}{\Phi},
\end{equation}
and in this work we only need its expression evaluated on flat spacetime that reads {\small 
\begin{align}
T_{\mu\nu} ={}& T_{\mu\nu}^\text{s,flat}
-\alpha \biggl\{ 4\frac{\sigma_{\mu\alpha}\sigma_\nu^{~\alpha}}{\sigma^2}
-2\biggl[2\frac{\sigma_{\mu\nu}}{\sigma}
-\eta_{\mu\nu} \biggl(\! \frac{\Box\sigma}{\sigma} - \frac{\sigma_\alpha \sigma^\alpha}{\sigma^2} \!\biggr)
\biggr]\nonumber \\
& \!\times\!\!\biggl(\! \frac{\Box\sigma}{\sigma} - \frac{\sigma_\beta \sigma^\beta}{\sigma^2} \!\biggr)\!
+ \!\eta_{\mu\nu} \biggl(\! 
\frac{(\sigma_\alpha \sigma^\alpha)^2}{\sigma^4} - 2\frac{\sigma_{\alpha\beta}\sigma^{\alpha\beta}}{\sigma^2} \!\biggr)\! \biggr\},\label{eq:T_munu}
\end{align}}%
where the indexes of $\sigma$ denote partial differentiation, and $T_{\mu\nu}^\text{s,flat}$ corresponds to \eqref{eq:T_munu_std} with $g_{\mu\nu}=\eta_{\mu\nu}$. Importantly, we remark that there is no contribution from the Weyl coupling, a conclusion that can be extrapolated to any conformally flat spacetime.

Finally, it is relevant to stress that the full action \eqref{eq:Non-NoethCSFaction} retains a global scaling symmetry when $\Omega=\text{const.}$ in \eqref{eq:CT}. Although the class of second-order theories allowing such property is much broader, see \cite{Padilla:2013jza} and more recently \cite{Babichev:2024krm}. An additional effective conformal symmetry appears for spacetimes with vanishing Gauss-Bonnet scalar, since the energy-momentum tensor again becomes traceless modulo the scalar equation of motion \cite{Fernandes:2021dsb}.

%%%%%%%%%%%%%%%%%%%%%%%%%%%%%%%%%%%%%%%%%%%%%%%%%%%%%%%%%%
\section{Noetherian conformal stealths
\label{sec:noeth-like}} 
%%%%%%%%%%%%%%%%%%%%%%%%%%%%%%%%%%%%%%%%%%%%%%%%%%%%%%%%%%

The Cheshire effect for the standard conformal scalar field \eqref{eq:stdCSF} was first shown to exist in Ref.~\cite{Ayon-Beato:2005yoq}. The related nonminimal conformal grin manifests through a flat spacetime stealth  parameterized as
\begin{equation}\label{eq:StealthNoeth_GeneralDep}
\Phi(x^\mu) = \frac{1}{ a x_\mu x^\mu + b_\mu x^\mu - c },
\end{equation}
where $x^\mu = (t, x^i)$ are the Cartesian coordinates of Minkowski spacetime and the integration constants $a$, $b_\mu$, and $c$ are tied in terms of the coupling constant of the conformal potential as
\begin{equation}
\lambda = 2ac + \frac12b_\mu b^\mu. \label{std_lambda}
\end{equation}
As proved in \cite{Ayon-Beato:2005yoq}, this is the general solution to the standard stealth constraints $T_{\mu\nu}^\text{s,flat}=0$, defining a \emph{Noetherian conformal stealth}. The above integration constants are determined up to isometries so they can be restricted further. Indeed, for non-vanishing $a$ one can fix $b_\mu = 0$ without any loss of generality by choosing an appropriate translation. In the same way, one can set $c = 0$ via a translation for vanishing $a$, while simultaneously aligning the vector $b_\mu$ through a rotation so that it has a single independent component.

Remarkably, the parameterization \eqref{eq:StealthNoeth_GeneralDep} is also effective  for the full non-Noetherian conformal action \eqref{eq:Non-NoethCSFaction}. This can be readily seen using the redefinition \eqref{eq:sigma}. The fact that the scalar field \eqref{eq:StealthNoeth_GeneralDep} prevails as flat spacetime stealth even when $\alpha\ne0$, relies in that the specific sum-separable form implied for $\sigma$ is characterized by having a Hessian proportional to the metric, $\sigma_{\mu\nu} = (\Box\sigma/4) \eta_{\mu\nu}$. This in turn forces the energy-momentum tensor \eqref{eq:T_munu} to be also proportional to the metric, $T_{\mu \nu} = (\text{tr} (T)/4) \eta_{\mu\nu}$, while the condition $\text{tr}(T) = 0$ amounts to the constraint 
\begin{equation}\label{eq:d_mp}
d_\mp \equiv \frac{1 \mp \sqrt{1-48\alpha\lambda}}{24\alpha}=2ac + \frac{1}{2} b_\mu b^\mu.
\end{equation}
For vanishing $\alpha$ only the upper-sign root is well defined giving the finite limit $d_- = \lambda$, recovering the known standard constraint \eqref{std_lambda}. The last configurations solve then the
non-Noetherian conformally invariant equation \eqref{eq:nonNoethCSFeom}. Since they observe the same local dependence than those of the case with conformally invariant action, and only the constraint obeyed by the integration constants is generalized, we shall refer to them as \emph{Noetherian-like conformal stealths}.

As we will show in the next section, the Noetherian-like conformal configurations above do not exhaust all the stealths allowed by the general action \eqref{eq:Non-NoethCSFaction}. Unfortunately, the goal of finding the general solution, as done in \cite{Ayon-Beato:2005yoq} for the $\alpha=0$ case, turns out to be a formidable task given that the stealth constraints that follow from \eqref{eq:T_munu} are no longer quasilinear. Hence, in what follows we will restrict ourselves to the spherically symmetric configurations, $\Phi = \Phi(t,r)$,  defined in flat spacetime 
\begin{equation}\label{eq:flatSpacetime}
ds^2 = -dt^2+dr^2+r^2d\Omega^2.
\end{equation}
For the Noetherian-like conformal configurations \eqref{eq:StealthNoeth_GeneralDep} this entails that $b_i = 0$, allowing only two different cases up to isometries, depending on whether $a$ is vanishing or not. For the former, one has the spatially homogeneous solutions 
\begin{equation}\label{eq:Noeth-homogeneous}
\Phi(t,r) = \frac{1}{\sqrt{ -2 d_\mp } \, t}.
\end{equation}
While the latter gives rise to the Lorentz-invariant family
\begin{equation} \label{eq:StealthNoeth}
\Phi(t,r) = \frac1{ a \left( -t^2 + r^2 \right) - d_\mp/(2a) },
\end{equation}
depending on the single integration constant $a$, which is a consequence of the scaling invariance of action \eqref{eq:Non-NoethCSFaction} discussed at the end of Sec.~\ref{sec:Non-Noeth-CSF}. For the special value of the non-Noetherian conformal coupling constant $\alpha = 1/(48\lambda)$, the two roots \eqref{eq:d_mp} degenerate to $d_\mp = 2 \lambda$ and one recovers the particular solutions first found in \cite{Babichev:2022awg}, cf.\ their Eqs.~(A8) and (A9). In fact, it was through these particular examples that they were the first to recognize that Noetherian conformal configurations can remain stealth in the presence of non-Noetherian conformal contributions. It is noteworthy that the non-Noetherian conformally supported flat universe reported in \cite{Fernandes:2021dsb} is just a conformal transformation of the stealth \eqref{eq:Noeth-homogeneous}. However, the transformed configuration is no longer free of backreaction like the original, due to the lack of conformal symmetry in the action.

The Noetherian and Noetherian-like conformal solutions studied in this section rely in the sum-separability of the inverse of the scalar field $\sigma$ \eqref{eq:sigma}. Fortunately,  when such separability is not attainable it is not the end of the story, at least for spherically symmetric configurations. In the next section, we show the existence of a new possibility, only valid in the non-Noetherian conformal regime $\alpha\ne0$.

%%%%%%%%%%%%%%%%%%%%%%%%%%%%%%%%%%%%%%%%%%%%%%%%%%%%%%%%%%%
\section{Non-Noetherian conformal stealths
\label{sec:strictly-nonNoeth}}
%%%%%%%%%%%%%%%%%%%%%%%%%%%%%%%%%%%%%%%%%%%%%%%%%%%%%%%%%%%

In this section, we will show that spherically symmetric generalized conformal stealths are necessarily divided into two branches. One corresponds to the separable sector already characterized in the previous section. Hence, we will concentrate in this section on the study of the new branch, which is radically different and its existence is only ensured for $\alpha\not=0$. Concretely, assuming spherical symmetry on conformal stealths gives a single off-diagonal component for the energy-momentum tensor \eqref{eq:T_munu}
\begin{equation}\label{eq:T_tr}
T_t^r=\frac{S\sigma_{tr}}{3\sigma^3}=0,
\end{equation}
where
\begin{equation}\label{eq:S}
S \equiv 1+12\alpha\left(\sigma_t^2-\sigma_r^2
+\frac2r \, \sigma\sigma_r\right).
\end{equation}
It is then evident that Noetherian conformal stealths, given by $\alpha = 0$, must be necessarily sum separable since the unique option is $\sigma_{tr}=0$. Interestingly, for $\alpha\ne0$ the vanishing of the quantity $S$ opens a new possibility that we will now explore. The remaining nontrivial stealth constraints can be conveniently rewritten from the energy-momentum tensor \eqref{eq:T_munu} as
\begin{subequations}\label{eq:stresst}
\begin{align}
T_t^t& =\frac{ 1-48\alpha\lambda
+\frac{16\alpha\sigma}{r}(\sigma_r-r\sigma_{rr})S-S^2 }{48\alpha\sigma^4}\!=0,\\
T_r^r& =\frac{ 1-48\alpha\lambda
+\frac{16\alpha\sigma}r(\sigma_r+r\sigma_{tt})S-S^2 }{48\alpha\sigma^4}=0,\\
T_\theta^\theta&=T_\phi^\phi= \frac{(r^2\sigma T_r^r)_r+r^2[\sigma_r
T_t^t-\sigma(T_t^r)_t]} {2r( \sigma-r\sigma_r )}=0.\label{eq:Tphi_phi}
\end{align}
\end{subequations}
The last expression is a consequence of the conservation of the energy-momentum tensor \eqref{eq:T_munu} and the fact that it is traceless in flat spacetime, since both conditions are satisfied via the scalar equation \eqref{eq:nonNoethCSFeom}. In other words, it is a by-product of two symmetries of action \eqref{eq:Non-NoethCSFaction}: general covariance and the already discussed effective conformal symmetry on flat spacetime.

One can then conclude from the previous equations that there are in fact new possibilities beyond the sum separability of the Noetherian-like conformal stealths determined by the conditions 
\begin{align}
&&S&=0, & \alpha\lambda &= \frac1{48}.&&
\label{eq:nonNoethBranch}
\end{align}
This is clearly a net departure from the standard stealths behavior, and an obvious manifestation of a Cheshire effect due to the non-Noetherian conformal contribution, since necessarily $\alpha\not=0$. We shall denote the resulting configurations as \emph{non-Noetherian conformal stealths}. From definition \eqref{eq:S} it follows that conditions \eqref{eq:nonNoethBranch} give rise to a nonlinear first-order PDE. In what follows we will show how to obtain its general solution using different but equivalent approaches. We start with an elementary and straightforward method to quasilinearize Hamilton-Jacobi equations that does not seem to have been considered in the PDE literature. 

%%%%%%%%%%%%%%%%%%%%%%%%%%%%%%%%%%%%%%%%%%%%%%%%%%%%%%%%%%
\subsection{Non-Noetherian from Hamilton-Jacobi 
quasilinearization \label{subsec:HJq}}
%%%%%%%%%%%%%%%%%%%%%%%%%%%%%%%%%%%%%%%%%%%%%%%%%%%%%%%%%%

If one rewrites the inverse scalar field \eqref{eq:sigma} as
\begin{equation}\label{eq:u}
\sigma = 2\sqrt{|\lambda|} \, r u,
\end{equation}
equations \eqref{eq:nonNoethBranch} with definition \eqref{eq:S} yield
\begin{equation}\label{eq:u_pde}
\frac{\epsilon S}{r^2}=u_t^2-u_r^2+\frac{u^2+\epsilon}{r^2}=0,
\end{equation}
where $\epsilon=\pm1$ is the sign of the non-Noetherian conformal coupling constant $\alpha=\epsilon|\alpha|$. Redefining again the involved function $u$ as
\begin{equation}\label{eq:u2Theta}
u=\begin{cases}
    \sinh\Theta, & \alpha>0, \\
    \cosh\Theta, & \alpha<0,
  \end{cases}
\end{equation}
we obtain in both cases the following Hamilton-Jacobi equation
\begin{equation}\label{eq:HJ}
\Theta_r^2-\Theta_t^2=\frac1{r^2}.
\end{equation}
This equation defines a hyperbola that can be locally parameterized by a function $H=H(t,r)$ as
\begin{align}
\Theta_r&=\frac{\cosh H}r, & \Theta_t&=\frac{\sinh H}r.
\label{eq:hyperb_param}
\end{align}
Such parameterization must be subject to the integrability condition $\partial_t\Theta_r=\partial_r\Theta_t$, that implies the function $H$ obeys the following inhomogeneous quasilinear first-order PDE
\begin{align}
X(H)&=\frac{\sinh H}r, & 
X&\equiv-\sinh H\partial_t+\cosh H\partial_r. \label{eq:qlfoPDE}
\end{align}
In other words, we have managed to transform the original fully nonlinear problem defined by Eq.~\eqref{eq:u_pde} to a quasilinear first-order PDE. The general solution to the latter can be found by integrating its characteristic system, see Eq.~\eqref{eq:CharSys_quasi}, which is
\begin{equation}\label{eq:CSqlfoPDE}
\frac{-dt}{\sinh H}=\frac{dr}{\cosh H}=\frac{r dH}{\sinh H}.
\end{equation}
The equation corresponding to the second equality is separable 
\begin{equation}
\frac{dr}{r}=\frac{d\sinh H}{\sinh H},
\end{equation}
which implies that the following function is a first integral of the system, namely
\begin{align}\label{eq:Omega}
\Omega &= \frac{\sinh H}{r}, & X(\Omega) &= 0.
\end{align}
As is custom in the treatment of quasilinear PDE, it is advantageous to use this integral as new coordinate
\begin{equation}\label{eq:from-tr-to-tOmega}
(t,r) \mapsto \left( t,\Omega=\frac{\sinh H(t,r)}r \right).
\end{equation}
Using these coordinates, the vector field becomes
\begin{equation}
X=X(t)\partial_t+X(\Omega)\partial_\Omega=-\sinh H\partial_t,
\end{equation}
and the PDE given in \eqref{eq:qlfoPDE} takes the simple form
\begin{equation}
\partial_t(\cosh H)=-\Omega.
\end{equation}
This allows its straightforward integration
\begin{equation}\label{eq:H(t,Omega)}
\cosh H(t,\Omega)=-\Omega t +\Omega h'(\Omega),
\end{equation}
depending on an arbitrary function of the invariant $\Omega$, appropriately chosen for later convenience. In terms of the original coordinates, this means that the most general function $H(t,r)$ satisfying the inhomogeneous quasilinear first-order PDE \eqref{eq:qlfoPDE} is implicitly given in terms of an arbitrary function by
\begin{equation}
t+r\coth H=h'\left(\frac{\sinh H}r\right).
\end{equation}
Equivalently, the most general function $\Omega(t,r)$ satisfying the homogeneous quasilinear first-order PDE involved in \eqref{eq:Omega} is given by the arbitrary implicit dependence
\begin{equation} \label{implicit-eq-for-Omega}
t+\frac1{\Omega}\sqrt{1+\Omega^2r^2}=h'(\Omega).
\end{equation}

Summarizing, the starting system \eqref{eq:hyperb_param} is only integrable for the found dependence of the local hyperbolic parameter $H$, expressed either in terms of the original or the adapted coordinates. This can be corroborated by writing the differential of the function $\Theta$ in terms of the new coordinates \eqref{eq:from-tr-to-tOmega}
\begin{align}
d\Theta&=\frac{\sinh H}r dt + \frac{\cosh H}r dr\nonumber\\
&=\Omega dt + \frac{\Omega\cosh H}{\sinh H}
d\left(\frac{\sinh H}{\Omega}\right),
\end{align}
or 
\begin{equation}
d\Theta - \frac{\cosh^2 H}{\sinh H}dH = \Omega dt - \frac{\cosh H}{\Omega}d\Omega.
\end{equation}
Using now the found dependence \eqref{eq:H(t,Omega)} in the right-hand side yields
\begin{equation}\label{eq:d(Theta...)=0}
d\left(\Theta - \cosh H - \ln\frac{\sinh H}{\cosh H + 1} - \Omega t + h \right)=0.
\end{equation}

We are now in a position to enunciate the main result of this subsection: \emph{the general solution to the Hamilton-Jacobi equation \eqref{eq:HJ} is given by}
\begin{subequations}\label{eq:generalTheta_H}
\begin{align}
\Theta(t,r)={}&\frac{t}r\sinh H + \cosh H 
- \text{arcsinh}\left( \frac1{\sinh H} \right) \nonumber\\
& - h\left(\frac{\sinh H}r\right),\\
0={}&t+r\coth H-h'\left(\frac{\sinh H}r\right),\label{eq:Himplicit}
\end{align}
\end{subequations}
\emph{where $h$ is an arbitrary single-argument function in terms of which the hyperbolic parameter determining the solution is implicitly defined as a local function $H(t,r)$ from the condition \eqref{eq:Himplicit}.} Notice that the integration constant appearing in \eqref{eq:d(Theta...)=0} can be absorbed into the function $h$, which is defined modulo a constant in \eqref{eq:H(t,Omega)}. An equivalent second statement is: \emph{the general solution to the Hamilton-Jacobi equation \eqref{eq:HJ} can be also expressed as}
\begin{subequations} \label{eq:generalTheta_Omega}
\begin{align}
\Theta(t,r) ={}& \Omega t + \sqrt{1+\Omega^2r^2}
- \text{arcsinh} \left( \frac1{\Omega r} \right) \nonumber\\
& - h(\Omega),\\
0 ={}& t+\frac{\sqrt{1+\Omega^2r^2}}{\Omega}-h'(\Omega),
\label{eq:Omegaimplicit}
\end{align}
\end{subequations}
\emph{where the solution is now determined by the function $\Omega(t,r)$, which is defined by the arbitrary implicit dependence implied from condition \eqref{eq:Omegaimplicit}.} This last expression for the general solution turns out to be just the one resulting from applying the procedure devised by Lagrange to the complete integral obtained from the Lagrange-Charpit method, as we will show in the next section.

Moreover, there exists a third possibility allowing an explicit expression for the general solution by using the integral $\Omega$ as a coordinate adapted to the problem. Concretely, using \eqref{eq:Omega}, \eqref{eq:H(t,Omega)} and \eqref{implicit-eq-for-Omega}, \emph{the general solution to the Hamilton-Jacobi equation \eqref{eq:HJ} also allows the explicit expression}
\begin{subequations}\label{eq:Theta(t,Omega)}
\begin{equation}
\Theta(t,\Omega) =\text{arcsinh}\!\left(\! 
\frac1{\Omega [ h'(\Omega) - t ]} \!\right)\! 
+ \Omega h'(\Omega) - h(\Omega),
\end{equation}
\emph{after changing the radial coordinate according to}
\begin{equation}
r = \frac{\sqrt{\Omega^2[h'(\Omega)-t]^2-1}}{\Omega}.
\label{eq:r2tOmega}
\end{equation}
\end{subequations}
The drawback of this explicit expression for the general solution is that in such coordinates the flat spacetime metric \eqref{eq:flatSpacetime} acquires a very unsuitable form.

Finally, the previous results allow to conclude that a non-Noetherian conformal Cheshire effect is possible thanks to the existence of a new $\alpha$-grin whose most general time-dependent spherically symmetric stealth is determined by
\begin{subequations}\label{eq:nonNCCeff}
\begin{align}
\Phi&=\begin{cases}
      \dfrac1{2\sqrt{|\lambda|}r\sinh\Theta}, & \quad \lambda>0, \\ \\
      \dfrac1{2\sqrt{|\lambda|}r\cosh\Theta}, & \quad \lambda<0,
      \end{cases}\\
\alpha&=\frac1{48\lambda},
\end{align}
\end{subequations}
where the function $\Theta$ allows either of the two implicit representations \eqref{eq:generalTheta_H} and \eqref{eq:generalTheta_Omega}, or even the explicit representation \eqref{eq:Theta(t,Omega)}.

%%%%%%%%%%%%%%%%%%%%%%%%%%%%%%%%%%%%%%%%%%%%%%%%%%%%%%%%%%
\subsection{Non-Noetherian from Lagrange-Charpit 
\label{subsec:GSfromLC}}
%%%%%%%%%%%%%%%%%%%%%%%%%%%%%%%%%%%%%%%%%%%%%%%%%%%%%%%%%%

It is also possible to tackle the full characterization of the spherically symmetric non-Noetherian conformal Cheshire effect by directly dealing with the PDE \eqref{eq:nonNoethBranch}. However, since this equation is no longer quasilinear as the obtained for $H$ in the previous subsection, the integration process leading to the general solution is less straightforward. It is then needed to appeal to the Lagrange approach reviewed in App.~\ref{app:Lagrange-approach}, where the general solution is constructed starting from a complete integral. Here we will apply such procedure to the complete integral for $u$ obtained using the Lagrange-Charpit method summarized in App.~\ref{app:LC_char}.

We concretely use the version (\ref{eq:u}-\ref{eq:u_pde}) of the nonlinear first-order PDE that can be written as
\begin{equation}\label{eq:F_PDE-for-u}
F(t,r,u,p,q) \equiv p^2-q^2+\frac{u^2+\epsilon}{r^2}=0,
\end{equation}
where we have defined $p=u_t$ and $q=u_r$. From Eq.~\eqref{eq:CharSys}, the
associated Lagrange-Charpit characteristic system reads {\small 
\begin{equation}\label{eq:CharSysnonNS}
\frac{\mathrm{d}t}{2p}=\frac{-\mathrm{d}r}{2q}=\frac{\mathrm{d}u}{2(p^2-q^2)}
=\frac{-r^2\mathrm{d}p}{2up}=\frac{-r^3\mathrm{d}q}{2(ruq-u^2-\epsilon)}.
\end{equation}}%
Using the original Eq.~\eqref{eq:F_PDE-for-u} in the third equality, we
obtain the separable equation
\begin{equation}\label{eq:third}
\frac{u\mathrm{d}u}{u^2+\epsilon} =\frac{\mathrm{d}p}{p},
\end{equation}
whose straightforward integration yields
\begin{equation}\label{eq:Hpde}
H(t,r,u,p,q)\equiv\frac{p}{\sqrt{u^2+\epsilon}}=c_1=\text{const.}
\end{equation}
Isolating the first derivatives from \eqref{eq:F_PDE-for-u} and
\eqref{eq:Hpde}, we obtain
\begin{equation}\label{eq:pqnonNS}
p=c_1\sqrt{u^2+\epsilon}, \qquad q=\frac{\sqrt{(u^2+\epsilon)
(1+{c_1\!}^2r^2)}}r,
\end{equation}
which allows to write $du=pdt+qdr$ as
\begin{equation}\label{eq:dunonNSint}
\frac{du}{\sqrt{u^2+\epsilon}}=c_1dt+\sqrt{1+{c_1\!}^2r^2}\frac{dr}r.
\end{equation}
The above Pfaff equation is integrable for the following potential {\small 
\begin{equation}\label{eq:d(.)=0}
d\biggl[\ln\!\left(\!\frac{\sqrt{1+{c_1\!}^2r^2}+1}{c_1r}
\bigl(u+\sqrt{u^2+\epsilon}\bigr)\!\!\right)\!
-c_1t-\sqrt{1+{c_1\!}^2r^2}\biggr]\!=0,
\end{equation}}%
from where the complete integral can be finally isolated using the redefinitions \eqref{eq:u2Theta} with {\small 
\begin{equation}\label{eq:CompIntLC}
\Theta(t,r;c_1,c_2)=c_1t + \sqrt{1+{c_1\!}^2r^2} 
- \text{arcsinh}\!\left(\! \frac1{c_1r} \!\right)\! - c_2.
\end{equation}}%

We now apply the Lagrange procedure to derive the general solution, starting from the above complete integral obtained via the Lagrange-Charpit method. 
Following App.~\ref{app:Lagrange-approach}, the first step is to impose in the complete integral \eqref{eq:u2Theta} with \eqref{eq:CompIntLC}, that $c_2 = h(c_1)$. Differentiating the result with respect to the remaining parameter, the next envelope condition is obtained 
\begin{equation}
\tilde u_1=\sqrt{u^2+\epsilon}\left(t+\frac{\sqrt{1+{c_1\!}^2r^2}}{c_1}-h'(c_1)\right)=0.
\end{equation}
The above condition can be explicitly solved for the parameter only for precise elections of the function $h$, defining its coordinate dependence $c_1=c_1(t,r)$. On the contrary, for an arbitrary function such dependence can only be defined implicitly. Consequently, according to the Lagrange procedure the general solution to the nonlinear first-order PDE \eqref{eq:F_PDE-for-u}, characterizing the spherically symmetric non-Noetherian conformal Cheshire effect, allows the following parametric representation in \eqref{eq:u2Theta}  {\small 
\begin{subequations}\label{eq:GS<-LC}
\begin{align}
\Theta(t,r) &= c_1t + \sqrt{1+{c_1\!}^2r^2}
- \text{arcsinh}\!\left(\! \frac1{c_1r} \!\right)\! 
- h(c_1),\label{eq:GS<-LC_I1}\\
0 &= t+\frac{\sqrt{1+{c_1\!}^2r^2}}{c_1}-h'(c_1),\label{eq:GS<-LC_I2}
\end{align}
\end{subequations}}%
where $c_1(t,r)$ is a function implicitly determined by the second condition. This is precisely the representation \eqref{eq:generalTheta_Omega} obtained in the previous subsection following the quasilinearization procedure and parameterized by the function $\Omega$, which is just the local dependence of $c_1$ in the present context. Consistently, both approaches give rise to the same representation of the general spherically symmetric non-Noetherian stealth configuration \eqref{eq:nonNCCeff}.

%%%%%%%%%%%%%%%%%%%%%%%%%%%%%%%%%%%%%%%%%%%%%%%%%%%%%%%%%%
\subsection{Non-Noetherian from Noetherian 
\label{subsec:GSnonNfromN}}
%%%%%%%%%%%%%%%%%%%%%%%%%%%%%%%%%%%%%%%%%%%%%%%%%%%%%%%%%%

As shown at the beginning of the section, conformal stealths with spherical symmetry in flat spacetime are divided into two branches: the Noetherian-like and the non-Noetherian ones. Surprisingly enough, it happens that the two classes of solutions intersect when $\alpha\lambda=1/48$. Indeed, it is known that other source of complete integrals for nonlinear first-order PDE are the separable solutions; in the case of \eqref{eq:nonNoethBranch} its most general sum-separable solution is
\begin{equation}\label{eq:sigmaNoeth}
\sigma(t,r;a,\tau) = a \left[ -( t-\tau )^2 + r^2 \right] - \frac{\lambda}{a},
\end{equation}
which is nothing other than the Noetherian-like conformal stealth \eqref{eq:StealthNoeth} at $\alpha=1/(48\lambda)$, before performing the time translation to make $b_0$ vanish in \eqref{eq:StealthNoeth_GeneralDep}. In other words, the Noetherian-like conformal stealths also provide a complete integral for the nonlinear first-order PDE \eqref{eq:nonNoethBranch} describing non-Noetherian conformal stealths!

With the different complete integral \eqref{eq:sigmaNoeth}, we can apply the Lagrange procedure again to obtain another parameterization of the general solution to the PDE \eqref{eq:nonNoethBranch}. Making that the original parameter arbitrarily depends on the reinstalled time-translation one, $a = a(\tau)$, and following App.~\ref{app:Lagrange-approach} again, it is obtained that the general solution can be also represented as
\begin{subequations}\label{eq:sigma-nonNoeth_a(tau)}
\begin{align}
\sigma(t,r) &= a(\tau) \left[ -( t-\tau )^2 + r^2 \right] - \frac{\lambda}{a(\tau)},\\
a'(\tau) &= -\frac{2 (t-\tau) a^3(\tau)}{\lambda + \left[ -( t-\tau )^2 + r^2 \right] a^2(\tau)},
\end{align}
\end{subequations}
where the function $\tau(t,r)$ is implicitly defined by the last condition.

The above representation of the general solution is linked to the one given in terms of $\Theta$ in \eqref{eq:generalTheta_Omega} in the following way. First, consider a Legendre transform of the arbitrary function of the representation \eqref{eq:generalTheta_Omega}
\begin{align}
Z(z)&=\Omega z - h(\Omega), &
z&=h'(\Omega), 
\end{align}
where, as usual, the last relation is used to solve for $\Omega=\Omega(z)$. Then the arbitrary function of the representation \eqref{eq:sigma-nonNoeth_a(tau)} and its argument are defined by the following map {\small 
\begin{equation}
(z,Z) \mapsto \biggl(
\tau = z - \frac{\tanh(Z)}{\Omega(z)} ,
 a = \sqrt{\lambda}\,\Omega(z) \cosh(Z) \biggr),
\end{equation}}%
when $\lambda > 0$. For $\lambda < 0$, the same map holds if $\lambda$ is replaced by its absolute value as well as $\sinh$ and $\cosh$ are interchanged in every occurrence.

%%%%%%%%%%%%%%%%%%%%%%%%%%%%%%%%%%%%%%%%%%%%%%%%%%%%%%%%%%
\subsection{Singular and particular solutions 
\label{subsec:S&Ps}}
%%%%%%%%%%%%%%%%%%%%%%%%%%%%%%%%%%%%%%%%%%%%%%%%%%%%%%%%%%

For $\lambda < 0$, there exists a solution, not contained in the general one, and called singular solution, cf.\ App.~\ref{app:envelopes_singular-sols}. Solving conditions \eqref{eq:singular-sol_system} for the equation \eqref{eq:nonNoethBranch}, using either its version \eqref{eq:S} or \eqref{eq:u_pde}, one obtains that the singular solution is given by
\begin{equation} \label{eq:stealth_static-singular}
\sigma(r) = \pm 2 \sqrt{-\lambda} \, r.
\end{equation}
The singular solution is not only spherically symmetric but also static, properties that are incompatible for the Noetherian-like stealths considered in Sec.~\ref{sec:noeth-like}.\footnote{It may seem like a linear radial dependence as \eqref{eq:stealth_static-singular} could cause a bad behavior on the angular stealth constraints \eqref{eq:Tphi_phi}. But this is not the case; it is merely an artifact of how those components are expressed in terms of the others.}

Let us now turn our attention to those solutions which do arise as particular cases of the general solution for spherically symmetric non-Noetherian conformal stealths. The first question to answer is if there are more static solutions. In fact, choosing $a(\tau)=A=\text{const.}$ in \eqref{eq:sigma-nonNoeth_a(tau)} gives $\tau = t$ and the solution becomes 
\begin{equation} \label{eq:stealth_static-quadratic}
\sigma(r) = A r^2 - \frac{\lambda}{A}.
\end{equation}
This is another one-parameter class of static solutions, which in contrast to the singular one is defined for any sign of $\lambda$, as long as it is related to $\alpha$ through \eqref{eq:nonNoethBranch}. Remarkably, the previous static stealths can be also obtained as the flat spacetime limit of the second branch of static and spherically symmetric non-Noetherian conformally dressed black holes found by Fernandes in \cite{Fernandes:2021dsb} precisely for $\alpha\lambda=1/48$. These configurations were originally derived in \cite{Babichev:2022awg} following such approach, cf.\ their Eq.~(A4). In contrast, the first branch of black holes given in \cite{Fernandes:2021dsb}, which exists for the different couplings relation $\alpha\lambda=1/144$, does not have a well-defined flat spacetime limit. That sheds some light on the relevance of the couplings special tuning \eqref{eq:nonNoethBranch} for the manifestation of a non-Noetherian conformal Cheshire effect. The previous two static solutions turn out to be the only ones within the spherically symmetric stealths giving rise to the conformal Cheshire effect. In fact, if one considers $\sigma = \sigma(r)$, the PDE \eqref{eq:nonNoethBranch} becomes a quadratic first-order ordinary equation whose two roots straightforwardly yield \eqref{eq:stealth_static-singular} and \eqref{eq:stealth_static-quadratic}. 

Another interesting particular solution is obtained if one chooses $a(\tau) = B/\tau$, for which \eqref{eq:sigma-nonNoeth_a(tau)} yields
\begin{equation}
\sigma(t,r) = 2 \left( B t \pm \sqrt{ (B^2+\lambda) (t^2-r^2) } \right).
\end{equation}
Spatially homogeneous stealths can be obtained from the previous one by setting $B = \pm\sqrt{-\lambda}$ for $\lambda < 0$. They coincide with the spatially-homogeneous Noetherian-like conformal stealths \eqref{eq:Noeth-homogeneous} when $\alpha\lambda=1/48$,  first obtained in \cite{Babichev:2022awg}. It is straightforward to check from PDE  \eqref{eq:nonNoethBranch} that these are the only allowed spatially homogeneous configurations. On the other hand, taking $B = 0$ a new Lorentz-invariant solution different from the Noetherian-like conformal stealth \eqref{eq:StealthNoeth} is obtained 
\begin{equation}
\sigma(t,r) = \pm 2\sqrt{ \lambda \left( t^2-r^2 \right)},
\end{equation}
valid inside or outside the light cone for positive or negative $\lambda$, respectively. By similar arguments to those used in the static case, the above solutions and  \eqref{eq:StealthNoeth} are the only Lorentz-invariant stealths with spherical symmetry related to the conformal Cheshire effect.

%%%%%%%%%%%%%%%%%%%%%%%%%%%%%%%%%%%%%%%%%%%%%%%%%%%%%%%%%%
\section{Conclusions \label{sec:conclu}}
%%%%%%%%%%%%%%%%%%%%%%%%%%%%%%%%%%%%%%%%%%%%%%%%%%%%%%%%%%

In this work, we have explicitly shown the existence of a non-Noetherian conformal Cheshire effect for scalar fields obeying the most general second-order conformally invariant equation derived from an action principle. 

We started by showing that the configurations supporting the original conformal Cheshire effect \cite{Ayon-Beato:2005yoq}, characterized by a sum separable inverse of the scalar field, are also  stealth solutions for the full theory. The original restriction of the separability constants in terms of the self-interaction coupling constant $\lambda$ was generalized to involve also the non-Noetherian coupling constant $\alpha$. Being the latter coupling the only one responsible for breaking the conformal invariance of the action, while preserving that of the scalar equation, we have dubbed such configurations Noetherian-like conformal stealths. 

Furthermore, we have shown that the presence of the $\alpha$-coupling is responsible for the manifestation of a new conformal Cheshire effect, which is strictly non-Noetherian in the sense that it only arises for $\alpha \neq 0$. We arrived to that conclusion by thoroughly studying the spherically symmetric stealths of the whole theory in flat spacetime. It results that the introduction of the $\alpha-$terms  allow a new branch of stealth solutions beyond the separability of the Noetherian-like conformal configurations. We managed to prove that this new sector is ruled by a single nonlinear first-order PDE. Additionally, it only occurs for the tuning $\alpha\lambda=1/48$ between the couplings, which in particular requires both to be non-vanishing. We have called the resulting configurations non-Noetherian conformal stealths. We fully characterized these latter using different but equivalent methods, with results related at most by a combination of Legendre and point transformations.

When a coordinate system conventionally adapted to the flat spacetime metric was used, implicit representations of the general solution were obtained as it is common in the study of nonlinear first-order PDE. Unexpectedly, an explicit representation was also obtained in a coordinate system adapted to the problem, but causing the flat metric to be represented in an unconventional manner

Another interesting result is that the generic Noetherian-like solutions are just the sum-separable sector of the nonlinear first-order PDE. An outstanding lesson is that, by avoiding a premature use of isometries, these solutions have enough integration constants to serve as a seed for generating the general solution describing the full non-Noetherian sector.

We explicitly show how to recover special solutions from the general picture, including the static ones, those that are spatially homogeneous, and finally the Lorentz-invariant examples. In particular, it results that there are only two static cases, and the remaining infinite family of new conformal stealths indeed overfly Minkowski spacetime without causing even the slightest backreaction.

It was pointed out that one of the particular non-Noetherian static stealths coincides with the one obtained in \cite{Babichev:2022awg} also for $\alpha\lambda=1/48$, as the flat spacetime limit of the static and spherically symmetric conformally dressed black hole found in \cite{Fernandes:2021dsb} for Einstein gravity coupled to the theory \eqref{eq:Non-NoethCSFaction} without Weyl coupling, i.e.\ $f=0$. The exhaustive analysis of this work unveils that this was just the tip of the iceberg, and that an infinite class of configurations exists for the same couplings restriction. 

Hence, the emergence of the same couplings relation both for the non-Noetherian conformal Cheshire effect and for non-Noetherian conformally dressed black hole solutions should not be considered as a mere accident, but rather due to some deeper physical reasons that deserve future exploration. For the moment, we have provided further evidence that specific nontrivial non-Noetherian conformal couplings are favored by allowing new interesting physical configurations that are inconceivable in the Noetherian conformal case we are used to.

As a final comment, it is important to emphasize that the configurations presented in this work may have physical relevance beyond the idealized flat spacetime scenario, particularly in the Newtonian regime. Since this regime represents a perturbation of flat spacetime that is not necessarily devoid of any matter, as considered  here, it must necessarily include the perturbation of the stealth field. In general, such a perturbation will break the stealth behavior of the scalar field yielding to a nontrivial contribution to the energy-momentum tensor, that will in turn lead to the coupling of all involved perturbations. Consequently, the perturbation of the original stealth field could influence the Newtonian potential. If such an effect were to become observable, it would undoubtedly have astrophysical consequences.

\begin{acknowledgments}
We would like to thank D.\ Flores-Alfonso, U.\ Hernandez-Vera, Julio A. M\'endez, T.\ Simon, D.\ Uribe, and J.~Zanelli for useful discussions and reviewing this work. Thanks are due also to an anonymous Referee for inquiring about the physical importance of the solutions. This work has been partially funded by Conahcyt grant A1-S-11548 and FONDECYT grant 1210889. P.A.S.\ work was supported by Conahcyt ``Estancias Posdoctorales por M\'exico'' contract I1200/320/2022.
\end{acknowledgments}

\appendix

%%%%%%%%%%%%%%%%%%%%%%%%%%%%%%%%%%%%%%%%%%%%%%%%%%%%%%%%%%
\section{Nonlinear first-order PDE 
\label{app:nonlinear-PDEs}}
%%%%%%%%%%%%%%%%%%%%%%%%%%%%%%%%%%%%%%%%%%%%%%%%%%%%%%%%%%

In order to make this work self-contained, in this appendix we will briefly summarize general methods to solve nonlinear first-order PDE heavily used in the main text. They can be found in the PDE literature \cite{Evans:2010,Courant:1962,Forsyth:1959}, see also \cite{Demidov:1982} and references therein for some historical remarks. 

We consider a general nonlinear first-order PDE
\begin{equation}\label{eq:F_generalPDE}
F(t, r, u, p, q)=0,
\end{equation}
where $p=u_t$ and $q=u_r$. There are different types of solutions, or integrals, allowed by the PDE. A \emph{general solution} is the one given in terms of an arbitrary single-argument function. Additionally, a \emph{complete integral} is a solution involving two arbitrary constants which cannot be combined into a single one.\footnote{The previous definitions are generalized to the case of $n$ independent variables in the following way: a general solution is one given in terms of an arbitrary function with $n-1$ arguments and a complete integral is one involving $n$ arbitrary constants and no less.} On the contrary, a solution that involves neither an arbitrary function nor arbitrary constants is called a \emph{particular solution} if it is contained as a particular case of a general or complete solution, and a \emph{singular solution} otherwise. The general and singular integrals can be obtained from a complete integral using \emph{envelopes}; an approach to eliminate constants and obtain other solutions. The standard procedure to find the general solution of a nonlinear first-order PDE is due to Lagrange, see \cite{Demidov:1982}, and consists in constructing the envelope resulting from assuming that one of the integration constants of a previously found complete integral is an arbitrary function of the other constants. This is why complete integrals are relevant in the context of nonlinear first-order PDE, and an established way to build them is the Lagrange-Charpit method. In the following sub-appendices we will review these methods to obtain complete, general, and singular solutions.

%%%%%%%%%%%%%%%%%%%%%%%%%%%%%%%%%%%%%%%%%%%%%%%%%%%%%%%%%%
\subsection{The Lagrange-Charpit method \label{app:LC_char}}
%%%%%%%%%%%%%%%%%%%%%%%%%%%%%%%%%%%%%%%%%%%%%%%%%%%%%%%%%%

The Lagrange-Charpit method is a systematic approach to determining complete integrals. The method to find the general solution of a quasilinear first-order PDE was originally given by Lagrange and rests in the integration of the associated characteristic ordinary system, see \eqref{eq:CharSys_quasi} below. Charpit introduced a generalized characteristic system for fully nonlinear first-order PDE, providing a method to find a complete integral. Such characteristic system is defined by
\begin{equation} \label{eq:CharSys}
\frac{\mathrm{d}t}{F_p}=\frac{\mathrm{d}r}{F_q}=\frac{\mathrm{d}u}{pF_p+qF_q}
=\frac{-\mathrm{d}p}{pF_u+F_t}=\frac{-\mathrm{d}q}{qF_u+F_r}.
\end{equation}
From this system, one needs to find a first integral independent of \eqref{eq:F_generalPDE}, let us say
\begin{equation}\label{eq:H}
H(t,r,u,p,q)=c_1=\text{const.}
\end{equation}
Then, isolating the first derivatives from both equations
\eqref{eq:F_generalPDE} and \eqref{eq:H} should give the explicit expressions
\begin{equation}\label{eq:pq}
p=p(t,r,u;c_1), \qquad q=q(t,r,u;c_1).
\end{equation}
The characteristic system \eqref{eq:CharSys} warrants that the one-form
\begin{equation}\label{eq:du}
du-p(t,r,u;c_1)dt-q(t,r,u;c_1)dr=0,
\end{equation}
satisfies its integrability conditions, then it is possible to straightforwardly integrate the first-order quasilinear PDE system
\begin{equation}\label{eq:qlsys}
u_t=p(t,r,u;c_1), \qquad u_r=q(t,r,u;c_1),
\end{equation}
to obtain the desired complete integral
\begin{equation}\label{eq:u_gen}
u=u(t,r;c_1,c_2).
\end{equation}

In the particular case in which the original PDE \eqref{eq:F_generalPDE} is quasilinear, namely given by
\begin{equation}\label{eq:quasi}
a_1(t,r,u) p + a_2(t,r,u) q = a_0(t,r,u),
\end{equation}
it is enough to consider only the first two equations of the characteristic system \eqref{eq:CharSys}, since they become the following subsystem for $t$, $r$, and $u$
\begin{equation}\label{eq:CharSys_quasi}
\frac{dt}{a_1} = \frac{dr}{a_2} = \frac{du}{a_0}.
\end{equation}
Correspondingly, the previous two equations allow for two first integrals $G(t,r,u)$ and $H(t,r,u)$, but the two-dimensional nature of \eqref{eq:quasi} implies that they are not independent. Hence, the general solution $u = u(t,r)$ is determined from
\begin{equation}
G = h(H),
\end{equation}
where $h$ is an arbitrary single-argument function. This is precisely the method used in Subsec.~\ref{subsec:HJq} to find the general solution for the spherically symmetric non-Noetherian conformal stealths after quasilinearizing the involved Hamilton-Jacobi equation.

%%%%%%%%%%%%%%%%%%%%%%%%%%%%%%%%%%%%%%%%%%%%%%%%%%%%%%%%%%
\subsection{Envelopes and singular solutions 
\label{app:envelopes_singular-sols}}
%%%%%%%%%%%%%%%%%%%%%%%%%%%%%%%%%%%%%%%%%%%%%%%%%%%%%%%%%%

Let us now consider a solution involving one or more parameters
\begin{equation}
u = u(x^\mu; c_i),
\end{equation}
and construct the following system by differentiating with respect to the parameters
\begin{equation}\label{eq:u_i}
u_i \equiv \partial_{c_i} u(x^\mu; c_i) = 0.
\end{equation}
Provided the above algebraic system can be solved for the parameters as functions of the independent variables, $c_i = c_i(x^\mu)$, the \emph{envelope} of the multiparametric solution is defined as the function 
\begin{equation}
v(x^\mu)=u(x^\mu; c_i(x^\mu)),
\end{equation}
and is also a solution by the conditions \eqref{eq:u_i}. 

It turns out that the envelope of a complete integral precisely yields a singular solution, when it exists. Alternatively, the singular solution can be obtained without appealing to a complete integral by eliminating $p$ and $q$ from the following system
\begin{align} \label{eq:singular-sol_system}
F &= 0, & F_p &= 0, & F_q &= 0.
\end{align}
These procedures are followed in Subsec.~\ref{subsec:S&Ps} to identify the singular solution involved in the problem. Envelopes are also essential in the Lagrange approach to find the general solution.

%%%%%%%%%%%%%%%%%%%%%%%%%%%%%%%%%%%%%%%%%%%%%%%%%%%%%%%%%%
\subsection{Lagrange approach to general solutions 
\label{app:Lagrange-approach}}
%%%%%%%%%%%%%%%%%%%%%%%%%%%%%%%%%%%%%%%%%%%%%%%%%%%%%%%%%%

Lagrange conceived the following approach in order to build the general solution of a nonlinear first-order PDE. Starting from a complete integral, $u = u(t, r; c_1, c_2)$, consider that one of its parameters is an arbitrary function of the other, let say $c_2=h(c_1)$. Then the general solution is the envelope, as defined in the previous sub-appendix, of the function
\begin{equation}
\hat{u} (t, r; c_1) \equiv u(t, r; c_1, h(c_1)).
\end{equation}
In terms of the resulting local dependence of the parameter, $c_1 = c_1(t,r)$, the general solution is formally expressed as
\begin{equation}
u_\text{gs} = u(t,r;c_1(t,r),h(c_1(t,r))),
\end{equation}
where is manifest the dependence on an arbitrary single-argument function. However, since the function $h$ is arbitrary, such envelope cannot be explicitly found in general. Hence, the solution is parameterized in terms of the function $c_1(t,r)$ which is determined implicitly by the envelope condition 
\begin{subequations}
\begin{align}
u &= u(t,r;c_1,h(c_1)),\\
0 &= u_1(t,r;c_1,h(c_1)) + u_2(t,r;c_1,h(c_1)) h'.
\end{align}
\end{subequations}
This is the approach we follow in Subsecs.~\ref{subsec:GSfromLC} and \ref{subsec:GSnonNfromN} to reach other parameterizations of the general spherically symmetric non-Noetherian conformal stealths.

%%%%%%%%%%%%%%%%%%%%%%%%%%%%%%%%%%%%%%%%%%%%%%%%%%%%%%%%%%
\section{Second-order conformally invariant scalars 
\label{app:R_GBtilde}}
%%%%%%%%%%%%%%%%%%%%%%%%%%%%%%%%%%%%%%%%%%%%%%%%%%%%%%%%%%

For completeness, we report here conformally invariant scalars built from the auxiliary metric \eqref{eq:aux-metric}, explicitly expressed in terms of the scalar field, $\Phi$, and the background metric, $g_{\mu\nu}$. The first corresponds to the auxiliary scalar curvature 
\begin{equation}
\tilde{R}=\frac1{\Phi^3}(R-6\Box)\Phi,\label{eq:Rtilde}\\
\end{equation}
and the second to the auxiliary Gauss-Bonnet scalar, which is compactly expressed as
\begin{subequations}\label{eq:GBtilde}
\begin{equation}
\tilde{\mathscr{G}}=\frac1{\Phi^4}\left( \mathscr{G}+\nabla_\mu J^\mu \right),\label{eq:G+J}
\end{equation}
where
\begin{align}
\frac18 J^\mu={}&\frac1{\Phi}G^{\mu\nu}\nabla_{\nu}\Phi
+\frac1{\Phi^2}(\nabla^\mu\Phi)\Box\Phi
-\frac1{\Phi^2}(\nabla_{\nu}\Phi)\nabla^{\mu}\nabla^{\nu}\Phi \nonumber\\
&+\frac1{\Phi^3}(\nabla^{\mu}\Phi)\nabla_{\nu}\Phi\nabla^{\nu}\Phi.
\label{eq:J}
\end{align}
\end{subequations}
Here $R$ and $\mathscr{G}$ are the analogous quantities defined from the metric $g_{\mu\nu}$, being $G^{\mu\nu}$ the Einstein tensor of the latter. Notice that compact expression \eqref{eq:G+J} is compatible with the topological nature of the Gauss-Bonnet terms. However, it has the inconvenience that seems to incorporate third-order contributions, but these actually cancel out when considering Bianchi and Ricci identities.

%%%%%%%%%%%%%%%%%%%%%%%%%%%%%%%%%

\bibliography{References}

\begin{thebibliography}{30}%
\makeatletter
\providecommand \@ifxundefined [1]{%
 \@ifx{#1\undefined}
}%
\providecommand \@ifnum [1]{%
 \ifnum #1\expandafter \@firstoftwo
 \else \expandafter \@secondoftwo
 \fi
}%
\providecommand \@ifx [1]{%
 \ifx #1\expandafter \@firstoftwo
 \else \expandafter \@secondoftwo
 \fi
}%
\providecommand \natexlab [1]{#1}%
\providecommand \enquote  [1]{``#1''}%
\providecommand \bibnamefont  [1]{#1}%
\providecommand \bibfnamefont [1]{#1}%
\providecommand \citenamefont [1]{#1}%
\providecommand \href@noop [0]{\@secondoftwo}%
\providecommand \href [0]{\begingroup \@sanitize@url \@href}%
\providecommand \@href[1]{\@@startlink{#1}\@@href}%
\providecommand \@@href[1]{\endgroup#1\@@endlink}%
\providecommand \@sanitize@url [0]{\catcode `\\12\catcode `\$12\catcode `\&12\catcode `\#12\catcode `\^12\catcode `\_12\catcode `\%12\relax}%
\providecommand \@@startlink[1]{}%
\providecommand \@@endlink[0]{}%
\providecommand \url  [0]{\begingroup\@sanitize@url \@url }%
\providecommand \@url [1]{\endgroup\@href {#1}{\urlprefix }}%
\providecommand \urlprefix  [0]{URL }%
\providecommand \Eprint [0]{\href }%
\providecommand \doibase [0]{https://doi.org/}%
\providecommand \selectlanguage [0]{\@gobble}%
\providecommand \bibinfo  [0]{\@secondoftwo}%
\providecommand \bibfield  [0]{\@secondoftwo}%
\providecommand \translation [1]{[#1]}%
\providecommand \BibitemOpen [0]{}%
\providecommand \bibitemStop [0]{}%
\providecommand \bibitemNoStop [0]{.\EOS\space}%
\providecommand \EOS [0]{\spacefactor3000\relax}%
\providecommand \BibitemShut  [1]{\csname bibitem#1\endcsname}%
\let\auto@bib@innerbib\@empty
%</preamble>
\bibitem [{\citenamefont {Ay\'on-Beato}\ \emph {et~al.}(2005{\natexlab{a}})\citenamefont {Ay\'on-Beato}, \citenamefont {Martinez}, \citenamefont {Troncoso},\ and\ \citenamefont {Zanelli}}]{Ayon-Beato:2005yoq}%
  \BibitemOpen
  \bibfield  {author} {\bibinfo {author} {\bibfnamefont {E.}~\bibnamefont {Ay\'on-Beato}}, \bibinfo {author} {\bibfnamefont {C.}~\bibnamefont {Martinez}}, \bibinfo {author} {\bibfnamefont {R.}~\bibnamefont {Troncoso}},\ and\ \bibinfo {author} {\bibfnamefont {J.}~\bibnamefont {Zanelli}},\ }\href {https://doi.org/10.1103/PhysRevD.71.104037} {\bibfield  {journal} {\bibinfo  {journal} {Phys. Rev. D}\ }\textbf {\bibinfo {volume} {71}},\ \bibinfo {pages} {104037} (\bibinfo {year} {2005}{\natexlab{a}})},\ \Eprint {https://arxiv.org/abs/hep-th/0505086} {arXiv:hep-th/0505086} \BibitemShut {NoStop}%
\bibitem [{\citenamefont {Ay\'on-Beato}\ \emph {et~al.}(2006)\citenamefont {Ay\'on-Beato}, \citenamefont {Martinez},\ and\ \citenamefont {Zanelli}}]{Ayon-Beato:2004nzi}%
  \BibitemOpen
  \bibfield  {author} {\bibinfo {author} {\bibfnamefont {E.}~\bibnamefont {Ay\'on-Beato}}, \bibinfo {author} {\bibfnamefont {C.}~\bibnamefont {Martinez}},\ and\ \bibinfo {author} {\bibfnamefont {J.}~\bibnamefont {Zanelli}},\ }\href {https://doi.org/10.1007/s10714-005-0213-x} {\bibfield  {journal} {\bibinfo  {journal} {Gen. Rel. Grav.}\ }\textbf {\bibinfo {volume} {38}},\ \bibinfo {pages} {145} (\bibinfo {year} {2006})},\ \Eprint {https://arxiv.org/abs/hep-th/0403228} {arXiv:hep-th/0403228} \BibitemShut {NoStop}%
\bibitem [{\citenamefont {Callan}\ \emph {et~al.}(1970)\citenamefont {Callan}, \citenamefont {Coleman},\ and\ \citenamefont {Jackiw}}]{Callan:1970ze}%
  \BibitemOpen
  \bibfield  {author} {\bibinfo {author} {\bibfnamefont {C.~G.}\ \bibnamefont {Callan}, \bibfnamefont {Jr.}}, \bibinfo {author} {\bibfnamefont {S.~R.}\ \bibnamefont {Coleman}},\ and\ \bibinfo {author} {\bibfnamefont {R.}~\bibnamefont {Jackiw}},\ }\href {https://doi.org/10.1016/0003-4916(70)90394-5} {\bibfield  {journal} {\bibinfo  {journal} {Annals Phys.}\ }\textbf {\bibinfo {volume} {59}},\ \bibinfo {pages} {42} (\bibinfo {year} {1970})}\BibitemShut {NoStop}%
\bibitem [{\citenamefont {Ay\'on-Beato}\ \emph {et~al.}(2005{\natexlab{b}})\citenamefont {Ay\'on-Beato}, \citenamefont {Martinez}, \citenamefont {Troncoso},\ and\ \citenamefont {Zanelli}}]{Ayon-Beato:SAdS}%
  \BibitemOpen
  \bibfield  {author} {\bibinfo {author} {\bibfnamefont {E.}~\bibnamefont {Ay\'on-Beato}}, \bibinfo {author} {\bibfnamefont {C.}~\bibnamefont {Martinez}}, \bibinfo {author} {\bibfnamefont {R.}~\bibnamefont {Troncoso}},\ and\ \bibinfo {author} {\bibfnamefont {J.}~\bibnamefont {Zanelli}},\ }\href@noop {} {\bibinfo {title} {{Stealths overflying (A)dS}}} (\bibinfo {year} {2005}{\natexlab{b}}),\ \bibinfo {note} {unpublished}\BibitemShut {NoStop}%
\bibitem [{\citenamefont {Ay\'on-Beato}\ \emph {et~al.}(2015)\citenamefont {Ay\'on-Beato}, \citenamefont {Hassa\"{\i}ne},\ and\ \citenamefont {Ju\'arez-Aubry}}]{Ayon-Beato:2015qfa}%
  \BibitemOpen
  \bibfield  {author} {\bibinfo {author} {\bibfnamefont {E.}~\bibnamefont {Ay\'on-Beato}}, \bibinfo {author} {\bibfnamefont {M.}~\bibnamefont {Hassa\"{\i}ne}},\ and\ \bibinfo {author} {\bibfnamefont {M.~M.}\ \bibnamefont {Ju\'arez-Aubry}},\ }\bibfield  {journal} {\bibinfo  {journal} {arXiv preprint}\ }\href {https://doi.org/10.48550/arXiv.1506.03545} {10.48550/arXiv.1506.03545} (\bibinfo {year} {2015})\BibitemShut {NoStop}%
\bibitem [{\citenamefont {Banerjee}\ \emph {et~al.}(2008)\citenamefont {Banerjee}, \citenamefont {Jain},\ and\ \citenamefont {Jatkar}}]{Banerjee:2006pr}%
  \BibitemOpen
  \bibfield  {author} {\bibinfo {author} {\bibfnamefont {N.}~\bibnamefont {Banerjee}}, \bibinfo {author} {\bibfnamefont {R.~K.}\ \bibnamefont {Jain}},\ and\ \bibinfo {author} {\bibfnamefont {D.~P.}\ \bibnamefont {Jatkar}},\ }\href {https://doi.org/10.1007/s10714-007-0516-1} {\bibfield  {journal} {\bibinfo  {journal} {Gen. Rel. Grav.}\ }\textbf {\bibinfo {volume} {40}},\ \bibinfo {pages} {93} (\bibinfo {year} {2008})},\ \Eprint {https://arxiv.org/abs/hep-th/0610109} {arXiv:hep-th/0610109} \BibitemShut {NoStop}%
\bibitem [{\citenamefont {Maeda}\ and\ \citenamefont {Maeda}(2012)}]{Maeda:2012tu}%
  \BibitemOpen
  \bibfield  {author} {\bibinfo {author} {\bibfnamefont {H.}~\bibnamefont {Maeda}}\ and\ \bibinfo {author} {\bibfnamefont {K.-i.}\ \bibnamefont {Maeda}},\ }\href {https://doi.org/10.1103/PhysRevD.86.124045} {\bibfield  {journal} {\bibinfo  {journal} {Phys. Rev. D}\ }\textbf {\bibinfo {volume} {86}},\ \bibinfo {pages} {124045} (\bibinfo {year} {2012})},\ \Eprint {https://arxiv.org/abs/1208.5777} {arXiv:1208.5777 [gr-qc]} \BibitemShut {NoStop}%
\bibitem [{\citenamefont {Ay\'on-Beato}\ \emph {et~al.}(2013)\citenamefont {Ay\'on-Beato}, \citenamefont {Garc\'\i{}a}, \citenamefont {Ram\'\i{}rez-Baca},\ and\ \citenamefont {Terrero-Escalante}}]{Ayon-Beato:2013bsa}%
  \BibitemOpen
  \bibfield  {author} {\bibinfo {author} {\bibfnamefont {E.}~\bibnamefont {Ay\'on-Beato}}, \bibinfo {author} {\bibfnamefont {A.~A.}\ \bibnamefont {Garc\'\i{}a}}, \bibinfo {author} {\bibfnamefont {P.~I.}\ \bibnamefont {Ram\'\i{}rez-Baca}},\ and\ \bibinfo {author} {\bibfnamefont {C.~A.}\ \bibnamefont {Terrero-Escalante}},\ }\href {https://doi.org/10.1103/PhysRevD.88.063523} {\bibfield  {journal} {\bibinfo  {journal} {Phys. Rev. D}\ }\textbf {\bibinfo {volume} {88}},\ \bibinfo {pages} {063523} (\bibinfo {year} {2013})},\ \Eprint {https://arxiv.org/abs/1307.6534} {arXiv:1307.6534 [gr-qc]} \BibitemShut {NoStop}%
\bibitem [{\citenamefont {Ay\'on-Beato}\ \emph {et~al.}(2018)\citenamefont {Ay\'on-Beato}, \citenamefont {Ram\'\i{}rez-Baca},\ and\ \citenamefont {Terrero-Escalante}}]{Ayon-Beato:2015mxf}%
  \BibitemOpen
  \bibfield  {author} {\bibinfo {author} {\bibfnamefont {E.}~\bibnamefont {Ay\'on-Beato}}, \bibinfo {author} {\bibfnamefont {P.~I.}\ \bibnamefont {Ram\'\i{}rez-Baca}},\ and\ \bibinfo {author} {\bibfnamefont {C.~A.}\ \bibnamefont {Terrero-Escalante}},\ }\href {https://doi.org/10.1103/PhysRevD.97.043505} {\bibfield  {journal} {\bibinfo  {journal} {Phys. Rev. D}\ }\textbf {\bibinfo {volume} {97}},\ \bibinfo {pages} {043505} (\bibinfo {year} {2018})},\ \Eprint {https://arxiv.org/abs/1512.09375} {arXiv:1512.09375 [gr-qc]} \BibitemShut {NoStop}%
\bibitem [{\citenamefont {Horndeski}(1974)}]{Horndeski:1974wa}%
  \BibitemOpen
  \bibfield  {author} {\bibinfo {author} {\bibfnamefont {G.~W.}\ \bibnamefont {Horndeski}},\ }\href {https://doi.org/10.1007/BF01807638} {\bibfield  {journal} {\bibinfo  {journal} {Int. J. Theor. Phys.}\ }\textbf {\bibinfo {volume} {10}},\ \bibinfo {pages} {363} (\bibinfo {year} {1974})}\BibitemShut {NoStop}%
\bibitem [{\citenamefont {Babichev}\ and\ \citenamefont {Charmousis}(2014)}]{Babichev:2013cya}%
  \BibitemOpen
  \bibfield  {author} {\bibinfo {author} {\bibfnamefont {E.}~\bibnamefont {Babichev}}\ and\ \bibinfo {author} {\bibfnamefont {C.}~\bibnamefont {Charmousis}},\ }\href {https://doi.org/10.1007/JHEP08(2014)106} {\bibfield  {journal} {\bibinfo  {journal} {J. High Energ. Phys.}\ }\textbf {\bibinfo {volume} {08}}\bibfield  {number} {\bibinfo  {number} { (2014)},\ \bibinfo {pages} {106}},\ }\Eprint {https://arxiv.org/abs/1312.3204} {arXiv:1312.3204 [gr-qc]} \BibitemShut {NoStop}%
\bibitem [{\citenamefont {Minamitsuji}\ and\ \citenamefont {Motohashi}(2018)}]{Minamitsuji:2018vuw}%
  \BibitemOpen
  \bibfield  {author} {\bibinfo {author} {\bibfnamefont {M.}~\bibnamefont {Minamitsuji}}\ and\ \bibinfo {author} {\bibfnamefont {H.}~\bibnamefont {Motohashi}},\ }\href {https://doi.org/10.1103/PhysRevD.98.084027} {\bibfield  {journal} {\bibinfo  {journal} {Phys. Rev. D}\ }\textbf {\bibinfo {volume} {98}},\ \bibinfo {pages} {084027} (\bibinfo {year} {2018})},\ \Eprint {https://arxiv.org/abs/1809.06611} {arXiv:1809.06611 [gr-qc]} \BibitemShut {NoStop}%
\bibitem [{\citenamefont {Giardino}\ \emph {et~al.}(2023)\citenamefont {Giardino}, \citenamefont {Giusti},\ and\ \citenamefont {Faraoni}}]{Giardino:2023qlu}%
  \BibitemOpen
  \bibfield  {author} {\bibinfo {author} {\bibfnamefont {S.}~\bibnamefont {Giardino}}, \bibinfo {author} {\bibfnamefont {A.}~\bibnamefont {Giusti}},\ and\ \bibinfo {author} {\bibfnamefont {V.}~\bibnamefont {Faraoni}},\ }\href {https://doi.org/10.1140/epjc/s10052-023-11697-3} {\bibfield  {journal} {\bibinfo  {journal} {Eur. Phys. J. C}\ }\textbf {\bibinfo {volume} {83}},\ \bibinfo {pages} {621} (\bibinfo {year} {2023})},\ \Eprint {https://arxiv.org/abs/2302.08550} {arXiv:2302.08550 [gr-qc]} \BibitemShut {NoStop}%
\bibitem [{\citenamefont {Gleyzes}\ \emph {et~al.}(2015)\citenamefont {Gleyzes}, \citenamefont {Langlois}, \citenamefont {Piazza},\ and\ \citenamefont {Vernizzi}}]{Gleyzes:2014dya}%
  \BibitemOpen
  \bibfield  {author} {\bibinfo {author} {\bibfnamefont {J.}~\bibnamefont {Gleyzes}}, \bibinfo {author} {\bibfnamefont {D.}~\bibnamefont {Langlois}}, \bibinfo {author} {\bibfnamefont {F.}~\bibnamefont {Piazza}},\ and\ \bibinfo {author} {\bibfnamefont {F.}~\bibnamefont {Vernizzi}},\ }\href {https://doi.org/10.1103/PhysRevLett.114.211101} {\bibfield  {journal} {\bibinfo  {journal} {Phys. Rev. Lett.}\ }\textbf {\bibinfo {volume} {114}},\ \bibinfo {pages} {211101} (\bibinfo {year} {2015})},\ \Eprint {https://arxiv.org/abs/1404.6495} {arXiv:1404.6495 [hep-th]} \BibitemShut {NoStop}%
\bibitem [{\citenamefont {Bakopoulos}\ \emph {et~al.}(2024)\citenamefont {Bakopoulos}, \citenamefont {Charmousis}, \citenamefont {Kanti}, \citenamefont {Lecoeur},\ and\ \citenamefont {Nakas}}]{Bakopoulos:2023fmv}%
  \BibitemOpen
  \bibfield  {author} {\bibinfo {author} {\bibfnamefont {A.}~\bibnamefont {Bakopoulos}}, \bibinfo {author} {\bibfnamefont {C.}~\bibnamefont {Charmousis}}, \bibinfo {author} {\bibfnamefont {P.}~\bibnamefont {Kanti}}, \bibinfo {author} {\bibfnamefont {N.}~\bibnamefont {Lecoeur}},\ and\ \bibinfo {author} {\bibfnamefont {T.}~\bibnamefont {Nakas}},\ }\href {https://doi.org/10.1103/PhysRevD.109.024032} {\bibfield  {journal} {\bibinfo  {journal} {Phys. Rev. D}\ }\textbf {\bibinfo {volume} {109}},\ \bibinfo {pages} {024032} (\bibinfo {year} {2024})},\ \Eprint {https://arxiv.org/abs/2310.11919} {arXiv:2310.11919 [gr-qc]} \BibitemShut {NoStop}%
\bibitem [{\citenamefont {Lecoeur}(2024)}]{Lecoeur:2024kwe}%
  \BibitemOpen
  \bibfield  {author} {\bibinfo {author} {\bibfnamefont {N.}~\bibnamefont {Lecoeur}},\ }\emph {\bibinfo {title} {{Exact black hole solutions in scalar-tensor theories}}},\ \href@noop {} {Ph.D. thesis},\ \bibinfo  {school} {Laboratoire de Physique des 2 Infinis Ir\`ene Joliot-Curie, France} (\bibinfo {year} {2024}),\ \Eprint {https://arxiv.org/abs/2406.11095} {arXiv:2406.11095 [gr-qc]} \BibitemShut {NoStop}%
\bibitem [{\citenamefont {Charmousis}\ \emph {et~al.}(2019)\citenamefont {Charmousis}, \citenamefont {Crisostomi}, \citenamefont {Gregory},\ and\ \citenamefont {Stergioulas}}]{Charmousis:2019vnf}%
  \BibitemOpen
  \bibfield  {author} {\bibinfo {author} {\bibfnamefont {C.}~\bibnamefont {Charmousis}}, \bibinfo {author} {\bibfnamefont {M.}~\bibnamefont {Crisostomi}}, \bibinfo {author} {\bibfnamefont {R.}~\bibnamefont {Gregory}},\ and\ \bibinfo {author} {\bibfnamefont {N.}~\bibnamefont {Stergioulas}},\ }\href {https://doi.org/10.1103/PhysRevD.100.084020} {\bibfield  {journal} {\bibinfo  {journal} {Phys. Rev. D}\ }\textbf {\bibinfo {volume} {100}},\ \bibinfo {pages} {084020} (\bibinfo {year} {2019})},\ \Eprint {https://arxiv.org/abs/1903.05519} {arXiv:1903.05519 [hep-th]} \BibitemShut {NoStop}%
\bibitem [{\citenamefont {Anson}\ \emph {et~al.}(2021)\citenamefont {Anson}, \citenamefont {Babichev}, \citenamefont {Charmousis},\ and\ \citenamefont {Hassaine}}]{Anson:2020trg}%
  \BibitemOpen
  \bibfield  {author} {\bibinfo {author} {\bibfnamefont {T.}~\bibnamefont {Anson}}, \bibinfo {author} {\bibfnamefont {E.}~\bibnamefont {Babichev}}, \bibinfo {author} {\bibfnamefont {C.}~\bibnamefont {Charmousis}},\ and\ \bibinfo {author} {\bibfnamefont {M.}~\bibnamefont {Hassaine}},\ }\href {https://doi.org/10.1007/JHEP01(2021)018} {\bibfield  {journal} {\bibinfo  {journal} {J. High Energ. Phys.}\ }\textbf {\bibinfo {volume} {01}}\bibfield  {number} {\bibinfo  {number} { (2021)},\ \bibinfo {pages} {018}},\ }\Eprint {https://arxiv.org/abs/2006.06461} {arXiv:2006.06461 [gr-qc]} \BibitemShut {NoStop}%
\bibitem [{\citenamefont {Ay\'on-Beato}\ and\ \citenamefont {Hassaine}(2024)}]{Ayon-Beato:2023bzp}%
  \BibitemOpen
  \bibfield  {author} {\bibinfo {author} {\bibfnamefont {E.}~\bibnamefont {Ay\'on-Beato}}\ and\ \bibinfo {author} {\bibfnamefont {M.}~\bibnamefont {Hassaine}},\ }\href {https://doi.org/10.1016/j.aop.2023.169567} {\bibfield  {journal} {\bibinfo  {journal} {Annals Phys.}\ }\textbf {\bibinfo {volume} {460}},\ \bibinfo {pages} {169567} (\bibinfo {year} {2024})},\ \Eprint {https://arxiv.org/abs/2305.09806} {arXiv:2305.09806 [hep-th]} \BibitemShut {NoStop}%
\bibitem [{\citenamefont {Fernandes}(2021)}]{Fernandes:2021dsb}%
  \BibitemOpen
  \bibfield  {author} {\bibinfo {author} {\bibfnamefont {P.~G.~S.}\ \bibnamefont {Fernandes}},\ }\href {https://doi.org/10.1103/PhysRevD.103.104065} {\bibfield  {journal} {\bibinfo  {journal} {Phys. Rev. D}\ }\textbf {\bibinfo {volume} {103}},\ \bibinfo {pages} {104065} (\bibinfo {year} {2021})},\ \Eprint {https://arxiv.org/abs/2105.04687} {arXiv:2105.04687 [gr-qc]} \BibitemShut {NoStop}%
\bibitem [{\citenamefont {Demidov}(1982)}]{Demidov:1982}%
  \BibitemOpen
  \bibfield  {author} {\bibinfo {author} {\bibfnamefont {S.~S.}\ \bibnamefont {Demidov}},\ }\href {https://doi.org/10.1007/bf00418753} {\bibfield  {journal} {\bibinfo  {journal} {Archive for History of Exact Sciences}\ }\textbf {\bibinfo {volume} {26}},\ \bibinfo {pages} {325} (\bibinfo {year} {1982})}\BibitemShut {NoStop}%
\bibitem [{\citenamefont {Landau}\ and\ \citenamefont {Lifshitz}(1960)}]{Landau:1960}%
  \BibitemOpen
  \bibfield  {author} {\bibinfo {author} {\bibfnamefont {L.~D.}\ \bibnamefont {Landau}}\ and\ \bibinfo {author} {\bibfnamefont {E.~M.}\ \bibnamefont {Lifshitz}},\ }\href {https://doi.org/10.1016/C2009-0-25569-3} {\emph {\bibinfo {title} {Mechanics}}}\ (\bibinfo  {publisher} {Pergamon Press},\ \bibinfo {year} {1960})\BibitemShut {NoStop}%
\bibitem [{\citenamefont {Jackiw}(2006)}]{Jackiw:2005su}%
  \BibitemOpen
  \bibfield  {author} {\bibinfo {author} {\bibfnamefont {R.}~\bibnamefont {Jackiw}},\ }\href {https://doi.org/10.1007/s11232-006-0090-9} {\bibfield  {journal} {\bibinfo  {journal} {Theor. Math. Phys.}\ }\textbf {\bibinfo {volume} {148}},\ \bibinfo {pages} {941} (\bibinfo {year} {2006})},\ \Eprint {https://arxiv.org/abs/hep-th/0511065} {arXiv:hep-th/0511065} \BibitemShut {NoStop}%
\bibitem [{\citenamefont {Ay\'on-Beato}\ and\ \citenamefont {Hassaine}(2023)}]{Ayon-Beato:2023lrn}%
  \BibitemOpen
  \bibfield  {author} {\bibinfo {author} {\bibfnamefont {E.}~\bibnamefont {Ay\'on-Beato}}\ and\ \bibinfo {author} {\bibfnamefont {M.}~\bibnamefont {Hassaine}},\ }\href {https://doi.org/10.1016/j.aop.2023.169446} {\bibfield  {journal} {\bibinfo  {journal} {Annals Phys.}\ }\textbf {\bibinfo {volume} {458}},\ \bibinfo {pages} {169446} (\bibinfo {year} {2023})},\ \Eprint {https://arxiv.org/abs/2307.04048} {arXiv:2307.04048 [hep-th]} \BibitemShut {NoStop}%
\bibitem [{\citenamefont {Padilla}\ \emph {et~al.}(2014)\citenamefont {Padilla}, \citenamefont {Stefanyszyn},\ and\ \citenamefont {Tsoukalas}}]{Padilla:2013jza}%
  \BibitemOpen
  \bibfield  {author} {\bibinfo {author} {\bibfnamefont {A.}~\bibnamefont {Padilla}}, \bibinfo {author} {\bibfnamefont {D.}~\bibnamefont {Stefanyszyn}},\ and\ \bibinfo {author} {\bibfnamefont {M.}~\bibnamefont {Tsoukalas}},\ }\href {https://doi.org/10.1103/PhysRevD.89.065009} {\bibfield  {journal} {\bibinfo  {journal} {Phys. Rev. D}\ }\textbf {\bibinfo {volume} {89}},\ \bibinfo {pages} {065009} (\bibinfo {year} {2014})},\ \Eprint {https://arxiv.org/abs/1312.0975} {arXiv:1312.0975 [hep-th]} \BibitemShut {NoStop}%
\bibitem [{\citenamefont {Babichev}\ \emph {et~al.}(2024)\citenamefont {Babichev}, \citenamefont {Charmousis}, \citenamefont {Hassaine},\ and\ \citenamefont {Lecoeur}}]{Babichev:2024krm}%
  \BibitemOpen
  \bibfield  {author} {\bibinfo {author} {\bibfnamefont {E.}~\bibnamefont {Babichev}}, \bibinfo {author} {\bibfnamefont {C.}~\bibnamefont {Charmousis}}, \bibinfo {author} {\bibfnamefont {M.}~\bibnamefont {Hassaine}},\ and\ \bibinfo {author} {\bibfnamefont {N.}~\bibnamefont {Lecoeur}},\ }\bibfield  {journal} {\bibinfo  {journal} {arXiv preprint}\ }\href {https://doi.org/10.1142/S0217751X24450039} {10.1142/S0217751X24450039} (\bibinfo {year} {2024})\BibitemShut {NoStop}%
\bibitem [{\citenamefont {Babichev}\ \emph {et~al.}(2022)\citenamefont {Babichev}, \citenamefont {Charmousis}, \citenamefont {Hassaine},\ and\ \citenamefont {Lecoeur}}]{Babichev:2022awg}%
  \BibitemOpen
  \bibfield  {author} {\bibinfo {author} {\bibfnamefont {E.}~\bibnamefont {Babichev}}, \bibinfo {author} {\bibfnamefont {C.}~\bibnamefont {Charmousis}}, \bibinfo {author} {\bibfnamefont {M.}~\bibnamefont {Hassaine}},\ and\ \bibinfo {author} {\bibfnamefont {N.}~\bibnamefont {Lecoeur}},\ }\href {https://doi.org/10.1103/PhysRevD.106.064039} {\bibfield  {journal} {\bibinfo  {journal} {Phys. Rev. D}\ }\textbf {\bibinfo {volume} {106}},\ \bibinfo {pages} {064039} (\bibinfo {year} {2022})},\ \Eprint {https://arxiv.org/abs/2206.11013} {arXiv:2206.11013 [gr-qc]} \BibitemShut {NoStop}%
\bibitem [{\citenamefont {Evans}(2010)}]{Evans:2010}%
  \BibitemOpen
  \bibfield  {author} {\bibinfo {author} {\bibfnamefont {L.~C.}\ \bibnamefont {Evans}},\ }\href {https://books.google.com.mx/books?id=Xnu0o_EJrCQC} {\emph {\bibinfo {title} {Partial Differential Equations}}}\ (\bibinfo  {publisher} {AMS},\ \bibinfo {year} {2010})\BibitemShut {NoStop}%
\bibitem [{\citenamefont {Courant}\ and\ \citenamefont {Hilbert}(1962)}]{Courant:1962}%
  \BibitemOpen
  \bibfield  {author} {\bibinfo {author} {\bibfnamefont {R.}~\bibnamefont {Courant}}\ and\ \bibinfo {author} {\bibfnamefont {D.}~\bibnamefont {Hilbert}},\ }\href {https://doi.org/10.1002/9783527617234} {\emph {\bibinfo {title} {Methods of Mathematical Physics}}},\ Vol.~\bibinfo {volume} {2}\ (\bibinfo  {publisher} {John Wiley and Sons},\ \bibinfo {year} {1962})\BibitemShut {NoStop}%
\bibitem [{\citenamefont {Forsyth}(1959)}]{Forsyth:1959}%
  \BibitemOpen
  \bibfield  {author} {\bibinfo {author} {\bibfnamefont {A.~R.}\ \bibnamefont {Forsyth}},\ }\href {https://books.google.com.mx/books?id=xjgVvgAACAAJ} {\emph {\bibinfo {title} {Theory of Differential Equations}}},\ Vol.~\bibinfo {volume} {V}\ (\bibinfo  {publisher} {Dover Publications},\ \bibinfo {year} {1959})\BibitemShut {NoStop}%
\end{thebibliography}%

\end{document}